  \documentstyle[12pt,aasms4]{article}

  \def\kms{{\rm\,km/s}}
  \def\msun{{\rm\,M_\odot}}

  \def\vol#1  {{{#1}{\rm,}\ }}
  
  \def\etal{et al.\ }

  \newcount\refno
  \newcount\rfno
  \def\eq{$^{\the\refno\ }$\advance\refno by 1}
  \def\ad{\advance\rfno by 1}

  \def\clock{\count0=\time \divide\count0 by 60
  \count1=\count0 \multiply\count1 by -60
 \advance\count1 by \time
           
 \number\count0:\ifnum\count1<10{0\number\count1}\else\number\count1\fi}

     \def\kms{\rm km\,s^{-1}}

\def\Gcm2{\rm G~cm^2}

\def\beq{\begin{equation}}
\def\eeq{\end{equation}}

\def\h2{${\rm\,H_2}$}

\def \date         {\ifcase\month \message{zero} \or
                    January \or February \or March \or
				  April \or May \or June 
                   \or July \or 
                         August \or September \or October
		  \or November \or 
									                      December \fi
                \space\number\day, \number\year}

\begin{document}
\title{Formation of First Stars Triggered by Collisions and Shockwaves: Prospect for High Star Formation Efficiency and High Ionizing Photon Escape Fraction}
\author{Renyue Cen\altaffilmark{1}}
\altaffiltext{1} {Princeton University Observatory,
		  Princeton University, Princeton, NJ 08544;
		  cen@astro.princeton.edu}

\begin{abstract}

We show that large, high-redshift ($z>10$)
galaxies with virial temperature in excess of $10^4$K
may be mostly comprised of cold atomic clouds 
which were formerly minihalos. 
These clouds move at a speed of $\sim 15-30\kms$
and collide with one another on a time scale of $10^7$yr.
The supersonic collisions result in spatially distributed
star formation with high efficiency.
We then show that most, subsequent star formation 
in cold atomic clouds
may be triggered by shockwaves launched
from the first stars formed in collisions.
Those shockwave-compressed clouds
are even more widespread 
spatially because of large imparted velocities
and some
can possibly escape into the intergalactic medium.
More importantly, with respect to cosmological reionization,
widespread star formation 
would allow a much higher ionizing photon escape fraction.
These favorable conditions may form the physical basis to 
enable the standard cosmological model to produce a 
reasonably high Thomson optical depth $\tau_e=0.10-0.14$.
A chain reaction of star formation 
in minihalos in the intergalactic space 
may be triggered by explosions in the intergalactic medium,
if minihalos are strongly clustered.
In this case, a still higher $\tau_e$ would be achievable.

\end{abstract}


\section{Introduction}

The Wilkinson Microwave Anisotropy Probe (WMAP)
polarization observations measured 
Thomson optical depth $\tau_e=0.17\pm 0.04$ ($68\%$) (Kogut \etal 2003),
requiring a full reionization to have occured
first at $z=20^{+10}_{-9}$, such as
in the double reionization picture (Cen 2003a).
This high Thomson optical depth is further confirmed
by a variety of combined data sets (Tegmark \etal 2003).
To ionize the universe at such a high redshift
demands both a high formation 
efficient ($\epsilon$) for Population III stars
and a large escape fraction ($f_{esc}$)
for the produced ionizing photons.
Quantitatively, one would require the product of the two
to be greater than $0.01$ 
in order to produce $\tau_e > 0.11$
(e.g., Cen 2003b; Haiman \& Holder 2003; 
Wyithe \& Loeb 2003; Ciardi, Ferrara \& White 2003; 
Ricotti \& Ostriker 2003)
to be consistent with the WMAP observations at $\sim 1.5\sigma$ level.
Conventional wisdom  
may view these assumptions somewhat too optimistic at first sight.

In this {\it Letter} we suggest that 
star formation at high redshift ($z>10$)
may occur in massive starbursts,
triggered by supersonic collisions 
between constituent cold atomic clouds
and blastwaves produced by explosions of first stars.
The consequences are (1) high star formation efficiency
and (2) high ionizing photon escape fraction.

\section{Collisions between Cold Atomic Gas Clouds in High Redshift Galaxies}

The slope of density power spectrum is $\sim -3$
in the cold dark matter cosmological models.
As a result, galaxies over a wide range of mass, 
$10^4-10^8\msun$ of interest here
at high redshift,
are expected to form nearly simultaneously,
when they form.
This would imply that 
large galaxies of mass $10^7-10^8\msun$
(virial temperature $T_v>10^4$K at $z>10$),
have most of their mass in small clouds,
which are formerly minihalos.
How much mass of a large galaxy is in minihalos?
This is computable approximately
using the Extended Press-Schechter (EPS) theory
(Bond \etal 1991;  Bower 1991; Lacey \& Cole 1993, LC hereafter).
We use the instantaneous merger rate 
for a halo of mass $M_p$ to merger with 
a halo of mass $\Delta M=M-M_p$ to form a halo
of mass $M$ (LC):
\begin{equation}
R_N(M_p\rightarrow M) dM = ({2\over \pi})^{1/2} {1\over t} |{d\ln\delta_c\over d\ln t}|
{1\over M}|{d\ln \sigma\over d\ln M}| {\delta_c(t)\over \sigma}
{1\over (1-\sigma^2/\sigma_p^2)^{3/2}} \\
\exp{[-{\delta_c(t)^2\over 2}({1\over \sigma^2}-{1\over\sigma_p^2})]} dM,
\end{equation}
\noindent
where 
$\sigma$ is the density variance of 
a sphere in the linear density field 
extrapolated to $z=0$ that contains mass $M$;
$\sigma_p$ is $\sigma$ for mass $M_p$;
$\delta_c(t)$ is the critical density for collapse (equation 2.1 of LC);
$t$ is age of the universe at the epoch examined.
The total mass accretion/merger rate of halos of mass
in the range $M_1$ to $M_2$ to form a halo of mass $M$ at $t$ is then
\begin{equation}
R_M(M,M1,M2) = \int_{M-M_2}^{M-M_1} R_N(M_p\rightarrow M)(M-M_p)dM_p.
\end{equation}
\noindent
Figure 1 shows 
the mass fraction of accreted objects
in the mass range $2\times 10^4-10^7\msun$
by a large halo with $M=1\times 10^8\msun$
as a function of redshift.
We see that $50-80\%$ of all mass 
in large galaxies with virial temperature greater than $10^4$K
at $z>10$ are in minihalos.
The standard WMAP-normalized cold dark matter model (Spergel \etal 2003)
is used: $\Omega_M=0.27$, $\Lambda=0.73$,
$\Omega_b=0.04$, $H_0=70$km/s/Mpc, $n=0.99$ and $\sigma_8=0.90$.

\begin{figure}
\plotone{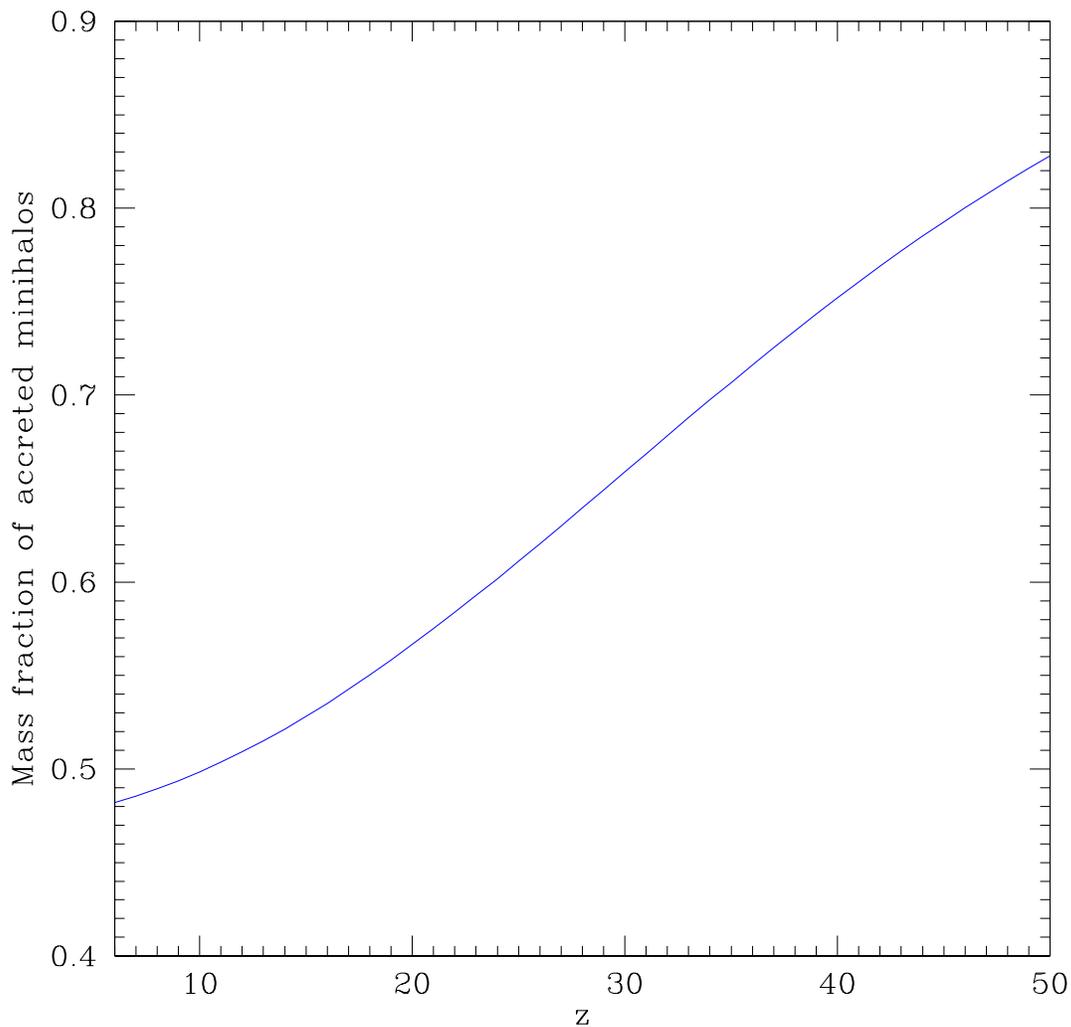}
\caption{
shows the fraction of total mass 
acquired by a halo of mass $M=1\times 10^8\msun$ at redshift $z$ 
that is in minihalos in the mass range $2\times 10^4-10^7\msun$.
Note $M=1\times 10^8\msun$ corresponds to a halo 
with a virial temperature of $3.7\times 10^4$K 
and halos with $M=2\times 10^4-10^7\msun$ have a virial temperature
of $120-7900$K, i.e., minihalos, at $z=20$. 
}
\label{y}
\end{figure}


The constituent minihalos do not remain intact in the process of infall and virialization.
First consider the effect due to gravitational tidal field
of the large, host halo on minihalos.
For simplicity we will assume that all dark matter halos are
singular isothermal spheres truncated at the respective virial radii,
$\rho(r) = \rho_v (r/r_v)^{-2}$ for $r\le r_v$,
where $\rho_v=100\bar\rho$, 
$\bar\rho$ is the mean density of the universe at that 
redshift and $r_v$ is the virial radius.
For illustration we will consider a large halo
of mass $1\times 10^8\msun$ at $z=20$ having a virial temperature
of $3.7\times 10^4$K.
We assume that $2/3$ of the total mass of the large halo
were originally from minihalos (see Figure 1),
and all minihalos have the same mass
of $1\times 10^6\msun$ with a virial temperature of $1700$K,
totaling $67$ such minihalos in the large halo.
We denote the virial radius of the large halo as $r_L$
($=0.71$kpc proper) 
and that of the minihalos as $r_M$ ($=0.15$kpc proper).
Then, a minihalo at radius $r$ from the center of the large halo
would have its matter exterior to 
$r_T=r_M (r/r_L)$ 
be tidally stripped off. 
The mean collision rate between minihalos 
(ignoring gravitational focus due to high speed) is 
\begin{equation}
<n\sigma v> = {\int_0^{r_L} v n_M(r) \sigma(r) n_M(r) 4\pi r^2 dr\over  \int_0^{r_L} n_M(r) 4\pi r^2 dr},  
\end{equation}
\noindent
where $\sigma(r)=\pi r_T^2$ is the cross section of a minihalo at
radius $r$;
$n_M(r)$ is the number density of minihalos at radius $r$;
$v$ is the velocity of each minihalo.
Assuming also an isothermal distribution
of minihalos, $n_M(r)=n_0(r/r_L)^{-2}$, we have
\begin{equation}
<n\sigma v> = \pi r_M^2 v n_0.
\end{equation}
\noindent
This result is obtained only considering the gravitational
tidal effect.
Let us now consider the effect due to external inter-minihalo medium (IMM)
on the internal gas of minihalos.
Consider a minihalo located at $r_L$.
The IMM with $T=3.7\times 10^4$K
and a density of $\rho_v/3$
has a pressure higher than the minihalo gas
with $T=1700$K and a density of $\rho_v$.
Hence, the gas density in minihalo would be raised 
by adiabatic compression by the external gas by a factor of $3.3$
and the temperature increases by a factor of $2.2$,
resulting in a pressure equilibrium 
(ignoring self-gravity of a minihalo for this simple calculation).
The compression would reduce the cross
section by a factor of $2.2$
(A better treatment will be using virial theorem with external
pressure but is not given here).
The percentage reduction in cross section 
at other radii is the same.
Thus, the average collision time between minihalos is 
\begin{equation}
t_C= {2.2\over \pi r_M^2 v n_0} = 4.3\times 10^7{\rm yr},
\end{equation}
\noindent
where the minihalo clouds in the large halo are assumed to 
move at a velocity equal to $v=\sqrt{3}\sigma_v=30\kms$
($\sigma_v=17.4\kms$ is 1-d the velocity dispersion of the large halo
for our illustrated case).
It is seen that the collision time among minihalos 
is about $25\%$ of the age of the Universe at $z=20$.
The time it takes for a minihalo to cross the 
large halo is $4.6\times 10^7$yr.
Thus, on average, a minihalo experiences a collision for every crossing time.

Gas within minihalos may be stripped off by ram pressure
when traveling through the IMM of the large host halo.
Applying the same physical argument 
of Gunn \& Gott (1972),
if the gravitational restoring force of a minihalo
is exceeded by ram pressure force, gas and dark matter
of the minihalo will be displaced, i.e., gas being stripped off.
This may be expressed by the condition
$\rho_{IMM} v_0^2 = \rho_M v_M^2$,
where $\rho_{IMM}$ is the density of the IMM;
$v_0$ is the traveling velocity of the minihalo;
$\rho_M$ is the mean density of a gas parcel of the minihalo in question;
$v_M$ is the 3-d velocity dispersion of the minihalo.
We see that this condition is similar to 
the pressure equilibrium condition above 
(without considering the self-gravity of minihalos).
Thus, ram pressure stripping is relevant
and likely to act at the same time when the cloud
is compressed by external thermal pressure.
It is possible that gas in outmost layers of a minihalo
may be subject to ram pressure stripping while
most interior gas will not be dislodged.
We will ignore the change to $t_C$ (equation 5) 
possibly caused by ram pressure stripping,
because other effects such as deformation of minihalo gas clouds
may easily dominate but are hard to characterize simply.

\begin{figure}
\plotone{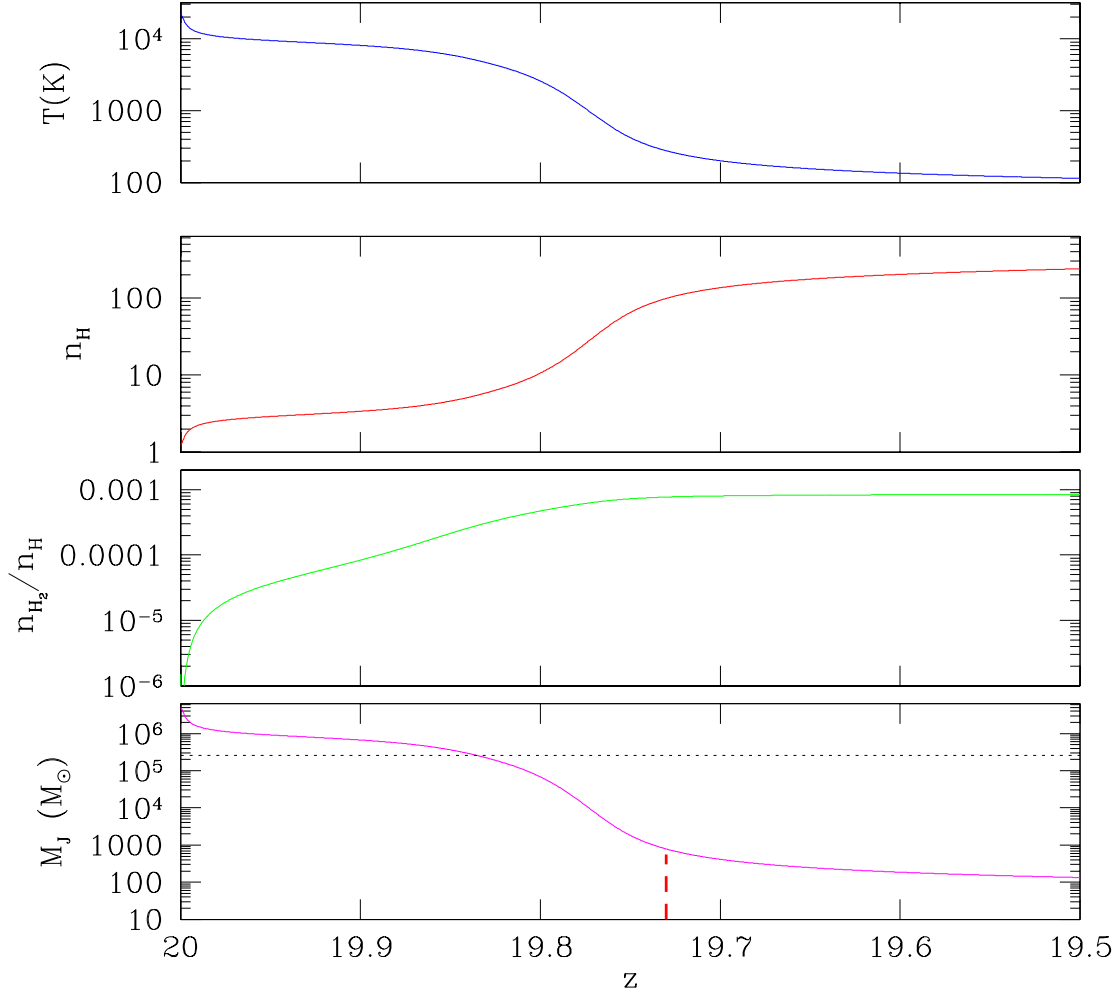}
\caption{
shows the evolution of the postshock gas
during a head-on collision between two cold atomic gas clouds.
The panels from top to bottom show the evolution of has
temperature, density, molecular hydrogen fraction and Jeans mass,
respectively.
Two cloud each has an initial velocity of $v_0=30\kms$ in the lab frame
with opposite direction with pre-collision parameters:
$T=1700$K and $n=0.42$cm$^{-3}$.
The vertical dashed tick sitting at the bottom axis of 
the bottom panel indicates
the elapsed time of $0.5t_s=3.64\times 10^6$yr, since the onset of the collision.
The black horizontal dotted line in the bottom panel indicates
the total baryonic mass of the collided clouds.
}
\label{y}
\end{figure}

What would happen in a head-on collision between two
clouds of cold gas?
Smith (1980) and Struck-Marcell (1982a,b) have  
studied this issue of cloud collision inside galaxies
for different ranges of Mach numbers using both analytic
method and simulations, in somewhat different environments.
Here we focus on the thermodynamic properties of post-shock gas,
computed assuming that post-shock region is isobaric,
while leaving detailed treatment to future
3-d hydrodynamic simulations.
All cooling/heating processes for a primordial gas
are incorporated, including atomic cooling, \h2 formation and cooling
and Compton cooling with the rates from Abel \etal (1997). 
Radiation background is assumed to exert no significant effect,
as shown by Mac Low \& Shull (1986) in the context of 
intergalactic shocks, which are physically similar to 
collision-induced shocks considered here.
Figure 2 shows the evolution of the postshock gas
following a head-on collision between two cold atomic gas clouds.
Note that the time for the shock to propagate through an entire cloud
is $t_s=2r_M/(4v_0/3)=7.3\times 10^6$yr.
Since the shock heated cloud has already cooled
to a much lower temperature 
and the Jeans mass has become much smaller
than the total baryonic mass of the merged cloud
at a time scale shorter than for the shock to transverse
the entire cloud,
the subsequent rarefaction wave following
the thermalization (shock) process
is insignificant, so are the compression and expansion phases
following that.
This indicates that the isobaric assumption 
is approximately valid and our simple calculation should 
have captured the essence of the process.
We see that a molecular hydrogen fraction of
$8\times 10^{-4}$ is reached and gas is cooled to
$\sim 100$K within a fraction of the Hubble time
after the collision ($\sim 10^6-10^7$yr).
The final Jeans mass is about $100\msun$,
indicating that stars formed may be massive.
One interesting difference in terms of star formation
between this picture and that of Abel \etal (2001)
is that multiple stars, possibly a star cluster,
may form, which, unlike in a single star case of Abel \etal (2001),
may provide an additional mechanism to remove
the angular momentum of the collapsing gas
through orbital coupling.

In summary, supersonic collisions between 
cold atomic gas clouds within large halos can trigger
star formation and star formation
efficiency may be quite high, because a large
portion of the colliding clouds may be able to cool
and form stars. The spatially distributed star formation
will be much more favorable to the escape of ionizing photons.

\section{Star Formation Triggered by Shockwaves}

The possible spread in time among collisions may be $10^7-10^8$yr,
longer than the lifetime of massive first stars of $\sim 3\times 10^6$yr.
Thus, it is likely that 
collisions between minihalo clouds
can not be considered coeval.
Consider two possible scenarios. 

In the first scenario, the first stars were so massive 
that they end their lives imploding and display no
explosive events, for example, if the stars
are more massive than $300\msun$ and non-rotating 
(Rakavy, Shaviv, \& Zinamon 1967;
Bond, Arnett, \& Carr 1984; 
Glatzel, Fricke, \& El Eid 1985;
Woosley 1986),
the subsequent star formation in collisions
that follow would proceed normally.
Star formation efficiency should be high
and so is the ionizing photon escape fraction,
since star formation will be spatially distributed
and extended (see below).

In the second scenario, first stars 
end in explosions and launch a blastwave.
Before computing the dynamics of blastwaves
from the explosion of first stars,
let us first consider the photoionization effect 
by the first stars on non-star-forming minihalos,
before blastwave encompassing them.
The ionizing photons 
from the massive stars formed in
collisions will propagate outward from the minihalo.
Since the ionization front (assuming an
ionizing photon emissivity of
$1.6\times 10^{48}~$photons~s$^{-1}$~$\msun^{-1}$; 
Bromm, Kudritzki, \& Loeb 2001) travels at a velocity of 
$v_{IF} = F/n = 3478 (d/1{\rm kpc})^{-2}(\epsilon/0.1) (n/0.1 {\rm cm}^{-3})^{-1}$km/s
(assuming no attenuation up to distance $d$ from the emitting source),
much faster than the sound speed in photoionized gas,
an R-type ionization front thus develops.
Subsequently, a sound wave (not a shockwave)
will develop driving the ionized gas
outward at the sound speed, which would travel
$0.046$kpc ($\sim r_M/3$) in the lifetime of the massive stars.
As a result, the gas density immediately
surrounding the stars will be much reduced.
Given that the expansion distance is comparable to the 
original minihalo radius one may expect
that the final gas distribution surrounding the stars at
the end of their lives may have a density comparable to mean density
of the original minihalo.
Without detailed simulations
we resort to some order of magnitude estimates.
Let us assume that the clumping factor of the gas in
the star forming minihalo cloud is $C$
(in terms of pre-shock gas density),
then it can be shown that the total amount of gas
required to balance the ionizing photon emission
from recombination is
$M_{rec} =1.0\times 10^5 (e/0.1) (C/16)^{-1}\msun$
for an ionizing photon luminosity
of $4.2\times 10^{52} (e/0.1)$sec$^{-1}$,
which should be compared to $2.6\times 10^5\msun$
of the total gas mass of the colliding clouds.
The fact that the former is of the same order of the latter
and the fact that the post-collision gas medium of the minihalos
is likely to be quite irregular
tell us that 
(1) a significant fraction of the ionizing photons is likely to escape 
the star-forming minihalos,
and 
(2) the Str\"omgren sphere size is not much larger the minihalo itself,
which would imply that the ionizing effect from the star-forming
minihalos on other minihalos in the same galaxy
may be relatively mild.
In a fraction of the solid angle, $f_\Omega$,
from the star-forming minihalo
optical depth may be small enough to allow 
for the ionizing photons to escape. 
For those directions, ignoring the recombination effect due to the IMM,
it can be shown that the minimal
distance from the star-forming minihalo
for the recipient minihalo of ionizing photons to
experience a D-type ionization front 
is $d_{min} = 2.2 ({\rho\over\rho_M})^{-1/2}$kpc,
which should be compared to the virial radius of
the large-host halo of $r_L=0.71$kpc.
Thus, within the fraction of the solid angle
where ionizing photons are able to escape into the IMM,
it appears that ionization front is R-type
initially at the virial radius of a minihalo
but turns into a D-type front when 
penetrating deeper into the minihalo
(at $0.57r_M$ for a minihalo at a distance of $r_L$).
A D-type front may transmit a weak shock 
causing slight compression of the gas,
while the regions cruised through
by the R-type front may expand outward.
In brief, we see that a small fraction of non-star-forming
minihalos within the solid angle of escaped ionizing photons
from star-forming minihalos
may experience some dynamic effect due to photoionization,
likely resulting in a very compact central region
and evaporated outskirts (Barkana \& Loeb 1999;
Shapiro \etal 2003).
Most of the minihalos are, however, not irradiated
by ionizing photons from the first stars 
and would remain neutral and intact.

Now consider the effects of 
the blastwaves produced by the explosions 
of the first stars 
in the same large host galaxy. 
We will  assume that the first stars each have
mass of $150\msun$ with supernova explosion energy of
$3\times 10^{52}$erg. 
A star formation efficiency of $\epsilon$ 
would give a total explosive energy 
of 
$5.2\times 10^{54}(\epsilon/0.1)$erg.
It is not required that
the mass of the first stars be in the range $130-300\msun$
to induce pair-instability supernovae;
our conclusions remain little changed if the mass of the first stars
is the range of $20-130\msun$.
We will now estimate how far and at what speed 
the blastwave from such an explosion would travel.
For simplicity and easy tractability
we assume that the blastwave has two phases:
the Sedov-Taylor phase followed by a momentum conserving
phase (with an instantaneous cooling phase in-between).
The assumed second phase is conservative in the sense that
it ignores any pressure of possibly still hot interior gas
on the shock shell.
The formula for the first phase is adopted
from Ostriker \& McKee (1988) assuming a uniform density locally.
Cooling rate for a primordial gas
is taken from Sutherland \& Dopita (1993)
with Compton cooling added.
Figure 3 shows the velocity of the blastwave after it has
traveled a distance 
such that 
it would, on average, hit $1$ (solid curve)
and $8$ (dashed curve) other minihalos,
as a function of the ambient gas density (depending upon the radius).
We see that for density in the relevant range
the expected shock velocity sweeping minihalos 
may be in the range $v_s=20-1000$km/s.
We note that, with significantly reduced density surrounding the exploding
stars due to photoionization, the blastwave
only starts cooling way after breaking out
the minihalo gas clouds in which the stars were formed.
Thus, neglecting the gas in the host minihalo does not introduce
significant errors.

\begin{figure}
\plotone{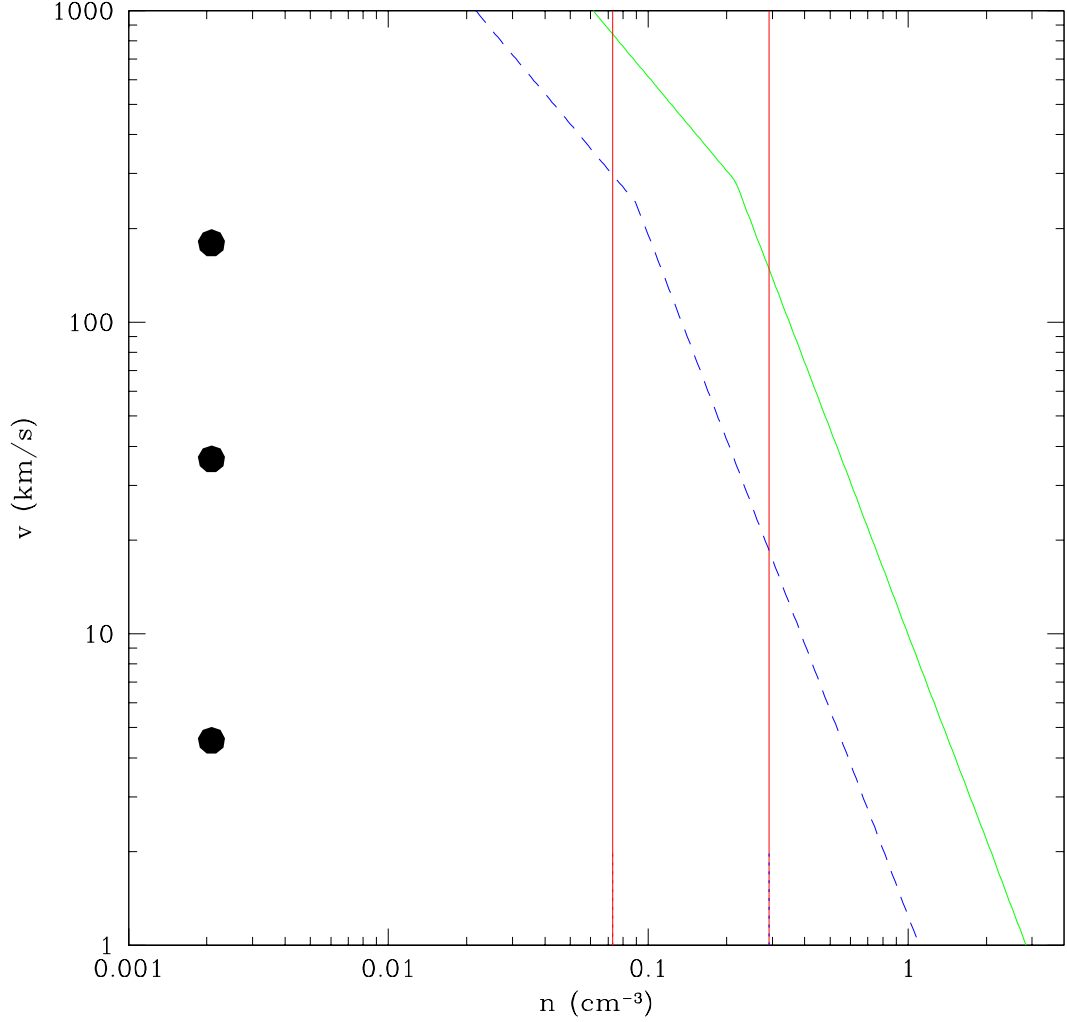}
\caption{
shows the velocity of the blastwave after it has
traveled a distance such that 
it would, on average, hit $1$ (solid curve)
and $8$ (dashed curve) other minihalos,
as a function of the ambient gas density (depending upon the radii).
The two vertical dotted lines 
enclose $50\%$ of the minihalos in the large host galaxy.
For later discussion (see below),
three solid dots indicate the velocity of the blastwave
from a minihalo with an explosion energy of 
$2.6\times 10^{54}$erg (for $\epsilon=0.1$)
after it has traveled a distance in the mean intergalactic medium 
such that it would have, on average, hit $0.016$, $0.13$ and $1$ 
(from top to bottom) other minihalos, respectively.
}
\label{y}
\end{figure}

What would happen to the minihalos when
shocks of this magnitude travel through them?
A shock with 
a velocity of $93$km/s in the ambient
medium would produce a shock 
of velocity $30$km/s ($=93(\rho_{IMM}/\rho_M)^{1/2}=93/\sqrt{9.3}$)
upon entering a minihalo,
producing results comparable to those shown in Figure 2.
One important difference is that, in this case of encompassing blastwaves,
the external thermal pressure is high
and could aid in compressing the shocked and cooled minihalo
gas clouds,
which will likely enhance star formation in the minihalos.
The reader may consult Kang \etal (1990) for a 1-d treatment
in the context of globular cluster formation
and Klein, McKee, \& Colella (1994) for a recent calculation.
In short, blastwaves from first massive stars
could cause star formation in other minihalos
in the same galaxy, yielding a widespread star formation.
Most likely, one generation of stars may be able to trigger star formation
in majority of the minihalos.

A higher Mach shock compresses
the minihalo gas to a higher density,
resulting in a lower Jeans mass (roughly inversely
proportional to the Mach number).
Thus, earlier collisions between the faster blastwave
and minihalos could lead to formation of lower mass stars,
even in metal-free gas.
We suggest that the observed, extremely low metallicity,
low mass star HE0107-5240 (Christlieb \etal 2002)
may be formed during some early time, high Mach shocked
minihalos, which are slightly contaminated with metals
(see Hideyuki \& Nomoto 2003).

One interesting consequence of sweeping blastwave is
that a large bulk velocity will be imparted
on the compressed minihalo clouds.
The expected bulk velocity in this case
is about $3v_s/4\sqrt{9.3}=0.25v_s$, where $v_s$ is the blastwave
velocity in the ambient gas and the factor 
$\sqrt{9.3}$ is due to the density ratio of
minihalo gas to the ambient gas.
Thus, these minihalos may subsequently move at
a velocity larger than 
the escape velocity of the large host halo ($v_{esc}=35\kms$),
if $v_s\ge 140\kms$.
Consequently, star formation may be widespread
spatially and it is even possible that some stars may
form after the minihalos have left the original host galaxy. 
One immediate implication is that ionizing photons
from these stars 
will be much easier to escape, giving
rise to escape fraction possibly well in excess
of $10\%$, whereas in the conventional
picture of central star formation where 
the escape fraction is estimated to be
below $1\%$ (Wood \& Loeb 2000; Ricotti \& Shull 2001).

\begin{figure}
\plotone{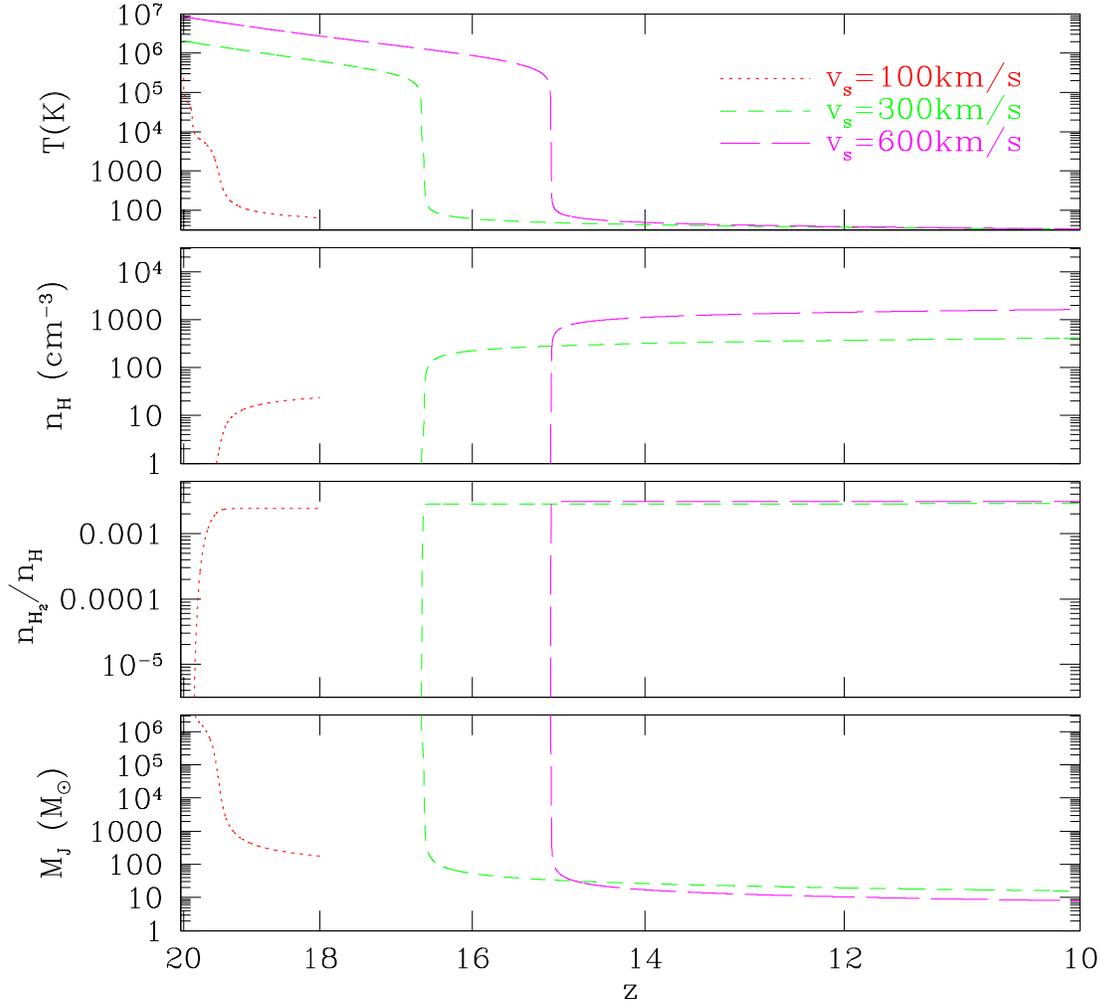}
\caption{
shows the evolution of the postshock gas
with three different shock velocities, $v_s=(100,300,600)$km/s,
respectively.
The panels from top to bottom show the evolution of has
temperature, density, molecular hydrogen fraction and Jeans mass,
respectively.
The shocks are assumed to propagate through a medium
with density equal to the mean gas density 
of the universe at $z=20$.
}
\label{y}
\end{figure}

Another implication is that, if some stars have
moved outside their original large host galaxy when
they explode,
the resulting blastwaves will travel
much further thanks to the lower density in the intergalactic
medium, which, in turn, could shock heat, accelerate and compress
minihalos in the intergalactic medium to form stars. 
If this chain reaction can be triggered,
star formation in minihalos, which are much more
abundant (in total mass) than large halos and are 
otherwise unable to form stars,
could produce many more ionizing photons
than what can be produced by large galaxies alone.
In Figure 3 the three solid dots 
show the velocity of the blastwave
from a minihalo with an explosion energy of 
$2.6\times 10^{54}$erg (for $\epsilon=0.1$)
after it has traveled a distance such that
it would have, {\it on average}, hit $0.016$, $0.13$ and $1$ 
other minihalos, respectively
(the fraction of mass in minihalos of mass $\sim 10^6\msun$
$f_M=10^{-3}$ is assumed).
If minihalos are strongly clustered
with a correlation length smaller than
$0.2d$ (where $d$ is the mean separation 
equal to $14(f/10^{-3})^{-1/3}$kpc proper
for $M_M=10^6\msun$ at $z=20$),
a chain reaction may be sustainable.
Minihalos may very well be strongly clustered
at high redshift (Barkana \& Loeb 2003).
We will study this separately.

In any event, 
blastwaves from stray stars propagating through the intergalactic medium
would create dense shells.
Figure 4 illustrate the temporal evolution of shocked gas 
at $z=20$ (assuming isobaric evolution);
preshock gas is at the mean gas density at $z=20$.
We see that cooling time is longer in this case
than shown in Figure 2;
here Compton cooling becomes important
and its cooling time scale (i.e, e-folding time) 
is $6\times 10^6$yr. 
On the longer time scales shown in Figure 4 
the detailed evolution would be affected
by concomitant dynamic evolution driven by 
gravitational instability of large-scale
density perturbations.
Further investigation is needed to 
see if star formation in shells occurs.
It is conceivable that a chain reaction may occur,
which would result in a picture analogous to the explosion
scenario proposed for larger galaxies
at lower redshift (Ostriker \& Cowie 1981).

\section{Conclusions}

In the cold dark matter model
large, high-redshift ($z>10$)
galaxies with virial temperature $T_v>10^4$K
are shown to be mostly comprised of cold atomic clouds,
which were originally minihalos and accreted into larger halos.
It is shown that 
supersonic collisions of cold atomic gas 
clouds on time scale of $10^7$yr
will shock heat and compress the gas, 
which then cools and form \h2 molecules
to further cool to $\sim 100$K.
Massive star formation ensues.
The resulting star formation is spatially distributed.

It is then shown that most of the star formation in minihalos
may be triggered by shockwaves produced by 
the first stars formed in collisions.
Those shockwave compressed minihalos 
will be more widespread 
spatially due to large imparted velocities
and some can possibly escape into the intergalactic medium.
If minihalos in the intergalactic space are strongly clustered,
a chain reaction of star formation 
in minihalos may be triggered initiated
by explosions of these stray stars.
In this case, far more ionizing photons can be produced
than large halos alone can.
Separately, whether cooling shells of swept-up
mean density gas in the intergalactic medium
can form stars and possibly trigger an explosion-induced
chain reaction even in 
the absence of strong minihalo clustering 
is unclear.

More importantly with respect to 
cosmological reionization,
widespread star formation 
would allow an ionizing photon escape fraction
much higher than in the conventional picture
where star formation occurs in the central
regions of galaxies
and whose the escape fraction is estimated to be below $1\%$
for galaxies more massive than $10^7\msun$.

Such star formation may be the norm
at $z>10$ and may create the physical basis to reconcile 
the standard cosmological model with 
the WMAP polarization observations.
An optical depth $\tau_e=0.10-0.14$ can be
achieved realistically even in the absence of chain
reactions forementioned.
A still higher optical depth may indicate
that some further enhancement in star formation
has occurred, such as through possible chain reactions.
Detailed simulations of star formation triggered by collisions
and blastwaves at high redshift galaxies 
would be required to firm up 
this picture and quantify key parameters,
including star formation efficiency and ionizing 
photon escape fraction.

Finally, we note that
these widespread star formation outlined above
may give rise to metal enrichment of the
intergalactic medium far more uniform than
any other processes that have been proposed
(e.g., Gnedin 1998; Aguirre \etal 2001; Cen \etal 2003).

This research is supported in part by grants AST-0206299 and NAG5-13381.

\end{document}